\documentclass[12pt]{article}
\usepackage[left=2.5cm,top=2.5cm,right=2.5cm,bottom=1.5cm]{geometry}
\usepackage{graphicx}
\usepackage{latexsym}
\usepackage{epstopdf}
\usepackage{amsmath}
\DeclareGraphicsExtensions{.eps}
\usepackage{cite}
\begin{document}
\begin{center}
\large{\bf{ Reconstruction of quintessence field  for the THDE  with swampland correspondence in $f(R,T)$ gravity}} \\
\vspace{10mm}
\normalsize{ Umesh Kumar Sharma }\\
\vspace{5mm}
\normalsize{Department of Mathematics, Institute of Applied Sciences \& Humanities, GLA University,
Mathura-281 406, Uttar Pradesh, India \\
\vspace{2mm}
E-mail: sharma.umesh@gla.ac.in \\}
\end{center}
\vspace{10mm}
\begin{abstract}
In the present work, we construct the Tsallis holographic quintessence model of dark energy in $f(R, T)$ gravity with Hubble horizon as IR cut-off. In a flat FRW background, the correspondence among the energy density of the quintessence model with the Tsallis holographic density permits the reconstruction of the dynamics and the potentials for the quintessence field.  The suggested Hubble horizon  infrared cut-off for the THDE density acts for two specific cases: (i) THDE 1 and (ii) THDE 2. We have reconstructed the Tsallis holographic quintessence model in the region $\omega_{\Lambda} > -1$ for the EoS parameter for both the cases. In addition, the quintessence phase of the THDE models is analyzed with swampland conjecture to describe the accelerated expansion of the Universe.\\

\end{abstract}
\smallskip
	Keywords: THDE; Quintessence; Swampland; $f(R, T)$ gravity. \\
\section{Introduction}

Gravity can thoroughly transform the dynamics of the physical system of the Universe, and that is described by invariance under spatial rotations and translations, at quite big scales. The appearance of the Universe is interesting because the expansion pace of the Universe has accelerated two times, first at extremely high energies and primordial times and the second succeeding period is enormously still ongoing (``Dark Energy'', DE). Whenever we try to discuss these wonders, it seems essential to reflect about the behavior of gravity itself. On referring the Universe as a whole in the elementary geometrical representation derived from classical GR (General Relativity), the need for summoning might indeed cast difficulties because the expansion of acceleration in Universe is two times.\\
 
  $\Lambda CDM$, the standard model of cosmology depends on GR and believes that all the components of DE, which are in the form of ordinary matter, cold dark matter (CDM) and a cosmological constant ($\Lambda$) created the Universe. Astonishingly, a representation of the Universe is provided through $\Lambda CDM$ and for this, all the data is compatible \cite{ref1, ref2}, the model exposes some lapses because of the problem of CC (cosmological constant). The obstacle starts from the mismatch among $\Lambda$'s value of the observed one, from the coincidence problem of late time and the one ordinarily expected from the corrections of quantum gravity \cite{ref3, ref4}. Moreover,  the model arises some sort of observational tensions between various datasets appear in the $\Lambda CDM$ model, for example, the linear power spectrum's amplitude at present times and the range of $8 h^{-1}Mpc$,  indicated through $\sigma_{8,0}$ and on the values of the Hubble constant $H_{0} (= 100 h kms^{-1} Mpc^{-1})$. 
 Comparing some examples as confined values of $H_{0}$ depends on the distance ladder of cosmic which points to a tension $4.4 \sigma$ \cite{ref6} and the radiation data of CMB (Cosmic Microwave Background) by Planck \cite{ref2} are all  at the level of background evolution of the Cosmos. In opposite SDSS (the Sloan Digital Sky Survey)\cite{ref10} and BAO (Baryon Acoustic Oscillations) analysis from BOSS (the Baryon Oscillation Spectroscopic Survey) \cite{ref11} presents inconsistency at $2.5 \sigma$ in $H_{0}$ by Planck \cite{ref12}. A discordance on $\sigma_{8,0}$ of around $2.3 \sigma$ could spot among Planck data  \cite{ref13, ref14} and KiDS (Kilo-Degree Survey)\cite{ref16} for perturbations of matter. The exploration for new physics beyond the standard model and the source of reflections on the gravity of $\Lambda CDM$, the ideas gathered from these surveys.\\
 
 The quintessence \cite{ref19}, a common scalar field with appropriate potential minimally linked to gravity and leads to inflation of late time is the next uncomplicated model recommended for DE. For the scalar field of spatially homogeneous quintessence,~ $\omega_{\Lambda} >-1$ is satisfied by the EoS and thus can present accelerated expansion. It avoids the difficulties with the initial conditions \cite{ref21} and also harmonious with its homogeneity, so this field is considered to be remarkably light. To illustrate the nature of DE except for quintessence many models of the scalar field have been introduced such as a scalar field with -ve kinetic energy, that present a solution named as phantom DE \cite{ref22}, a scalar field model with the non-standard kinetic term \cite{ref23, ref24} known as $k$-essence, dilation \cite{ref25} and tachyon \cite{ref26}. The modified theories of gravity known as $f(R)$ gravity, where DE arises by the change of the geometrical part \cite{ref27, ref28}, interacting DE models \cite{ref31, ref32}, DE models involving non-standard EoS \cite{ref33, ref34} and brane-world models \cite{ref35, ref36}. Other recommended models which include DE (see \cite{ref21} to review of all mentioned and other DE). Although the potential is chosen as support by phenomenological thoughtfulness in all mentioned scalar field models, which require the theoretical introduction. Some data linked to the theory of quantum gravity, a framework of a fundamental theory is another solution of the DE problem identified as a principle of holographic \cite{ref37, ref40, ref41} and generalized entropy formalism \cite{ref42, ref44, ref45}.\\
 
 Recently, four new holographic dark energy models (HDE) \cite{ref45a,ref45b, ref46, ref47} have been introduced to associate multiple generalized entropies of the FRW Universe horizon. In the present work, the focus is on the Tsallis holographic dark energy \cite{ref45a} inspired by the Tsallis entropy \cite{ref42}. The point that practicing system horizon in the Tsallis statistics, we can achieve the Bekenstein entropy, which is the determination of all these efforts \cite{ref48, ref51}. The authors also exposed adequate stability by the concerned models \cite{ref45a,ref45b, ref46, ref47}. To examine the cosmological and gravitational systems of the generalized statistical mechanics \cite{ref67, ref68, ref69, ref72, ref73, ref75, ref77, ref78} framework, positively a crucial role is performed by Tsallis definition of entropy \cite{ref42}.  Generally, the study of quantum gravity  \cite{ref88} confirms the conclusion that the content of the Tsallis entropy system is a power-law function of the system’s field \cite{ref42}.\\
 
 Reminding the derivation and definition of the energy density  $\rho_{\Lambda} = 3 c^{2} m^{2}_{p} / L^{2}$  of standard holographic depends on the connection  $ S \sim A \sim L^{2}$ of entropy area of black holes, here $ A = 4 \pi L^{2}$ presents the horizon's area \cite{ref40}. Due to the consideration of quantum \cite{ref89, ref90}, we get the modified HDE definition. In \cite{ref42}, Tsallis and Citro have exposed that we can modify the black hole's entropy of horizon as: $S_{\delta} = \gamma A^{\delta}$, where $\delta$ indicates the non-additivity parameter \cite{ref42} and $\gamma$ is an unknown constant. For the suitable limit of $\gamma = 1/4G $ and $\delta=1$  (in the link here $c = h = k_{B} = 1$), the Bekenstein entropy is obtained. Particularly, the common probability of probability allocation \cite{ref42} described, the system at this point or the power-law of a probability distribution must have shifted to worthless. The quantum gravity has confirmed the association \cite{ref88} and have shown the impressive results in the holographic and cosmological setups \cite{ ref75, ref77}.
The holographic principle states that the scaling of the freedom's physical system of the number of degrees should exhibit through its bounding area rather than its volume \cite{ref37, ref41}, and the infrared cutoff should compel it, a connection among UV ($\Lambda$) and the system entropy (S) and the IR (L) cutoffs is proposed by cohen et al. \cite{ref40} as
\begin{eqnarray}
\label{eq2}
L^{3} \Lambda^{3} \leq (S)^{3/4},
\end{eqnarray}
after combining with  $S_{\delta} = \gamma A^{\delta}$, it heads to \cite{ref40}, $\Lambda^{4} \leq (\gamma(4\pi)^{\delta})L^{2\delta-4}$, here $\Lambda^{4}$ is the energy density of the vacuum, in the hypothesis of HDE, it is DE's energy density ($\rho_{\Lambda}$) \cite{ref42, ref91}. Practicing this inequality the THDE density can be written as 
\begin{eqnarray}
\label{eq4}
\rho_{\Lambda} = BL^{2\delta-4},
\end{eqnarray}
here $B$ is an unknown parameter \cite{ref42, ref47}. Considering the Hubble horizon as a flat FRW Universe and for IR cutoff, a suitable applicant, besides no interaction among the other elements of cosmos and the candidate of DE. Hence, the law of conservation and the energy density matches to THDE in this way ($L = H^{-1}$),                     

\begin{eqnarray}
\label{eq5}
\rho_{\Lambda}=BH^{-2\delta+4}.
\end{eqnarray}

For the HDE model \cite{ref93}, the authors presented a new infrared cut off named GO horizon cutoff, they have also represented the correspondence among the quintessence, dilation, tachyon, and $k$-essence energy density in the flat FRW Universe for HDE density. Toward accelerated expansion, the correspondence also allows the dynamics and the potentials to reconstruct the model of the scalar field. The researcher in \cite{ref94} examined the conformity among the DBI-essence, tachyon, and the quintessence scalar field models in the core of chameleon Brans-Dicke cosmology with the model of NHDE. They have rebuilt the dynamics and the potentials for the model of the scalar field in the connection of chameleon Brans-Dicke cosmology by concluding the appearance of Hubble parameter and energy density. They remarked that with the evolution of the Universe, the potential increases as the matter-chameleon coupling becomes stronger. The conformity among the Chaplygin gas model, $k$-essence, the quintessence, dilation, and tachyon scalar field is shown in \cite{ref95}, in the non-flat Brans-Dicke Universe with the new HDE model. For this model, they reconstructed the dynamics and the potentials. In the Chaplygin gas and new holographic quintessence model for accelerated expansion in Brans-Dicke cosmology, they have met some limitations to the related parameter of the model to the potentials. They have shown new results in the frame of Brans-Dicke cosmology and also introduced special cases of new HDE in Einstein gravity. The researcher in \cite{ref96} examined the generalization of the conformity between tachyon, dilation, $k$-essence, and the quintessence with the new HDE model in the Universe of non-flat FRW.  They illustrated the accelerated expansion by reconstructing the dynamics and the potentials of these scalar field models. In the frame of Brans-Dicke cosmology, \cite{ref97} inquired about the cosmological relationship of interacting holographic energy density. They achieved the deceleration and the EoS parameter of HDE in the Universe of non-flat. They combined HDE and Brans-Dicke field to support $\omega_{\Lambda} = -1$ loop for the EoS of non-interacting HDE. They formed Einstein field equations to transform $\omega_{\Lambda}$ to the regime of the phantom when the interaction among dark matter and dark energy was carried. In \cite{ref98}, for the Universe of non-flat FRW in the interaction of the new age graphic DE model, the correspondence among dilation, $k$-essence, and tachyon is studied.  For the expansion of acceleration in the Universe, they replaced the dynamics and the potentials for the scalar field model. \cite{ref99} investigated the correspondence in the Einstein gravity framework among $k$-essence, dilation, and tachyon scalar field models with the interaction of the viscous ghost DE model. They reflected the Universe of non-flat FRW with interacting viscous ghost dark matter and dark energy. According to the evolutionary performance of the interacting viscous ghost DE model, for scalar field models, they reformed the potentials and the dynamics that characterized the Universe's accelerated expansion.\\

It is exciting to consider how the model of the scalar field can characterize the HDE as useful theories if we see the holographic vacuum energy situation as tending towards the underlying DE theory and observing an adequate explanation of the underlying DE theory for the models of the scalar field.
 \cite{ref100, ref101} reviewed the Chaplygin gas and the holographic phantom quintessence, \cite{ref102, ref103} presented the holographic tachyon models. All these models used the infrared cut off for future event horizon to reformed the conformity of the potentials. Though \cite{ref104} shows the conflict in cut-offs. In \cite{ref105}, they achieved other techniques of reconstruction in theories with multiple or single scalar fields. Here in this work, I am interested to represent the Tsallis holographic scenario of quintessence with an infrared cut off shown in Eq. (5) for the scalar field models as proposed in \cite{ref105a,ref106} for the holographic DE. As we retaliate the constants $\delta$ and $B$, we can reconstruct the potentials for the quintessence field for the THDE in $f(R, T)$ gravity.\\

The rest of the paper is organized as: we briefly review the $f(R, T)$ gravity in Sec. $2$.  The cosmological model and the THDE EoS parameter are discussed in Sec. $3$.  In Sec. $4$, we construct the correspondence between the THDE model and the quintessence field.  In Sec. $5$, we explain the correspondence with the Swampland criteria of the THDE model. At last in Sect. $6$, we finish up our discussion.\\

\section{Review on Gravitational field equations of $f(R, T)$ gravity}

Prior to showing the present contribution taking into account the work referenced, let us have a concise outline of the $F(R, T)$ gravity as the present work is going to investigate a cosmological reconstruction in the structure of the $F(R, T)$ gravity. The theory proposes a revised gravity work performed through the action given in \cite{ref107}

\begin{eqnarray}
\label{eq6}
S=\int{L_{m} \sqrt{-g}d^{4}x} +\frac{1}{16\pi}\int{f(R,T)\sqrt{-g}d^{4}x} ,
\end{eqnarray}

where, $L_{m}$ is Lagrangian matter density. Also,  $R$ represents the  ricci scalar and $T$ represents the trace of the matter energy momentum tensor ($T_{\mu \nu}$) for the arbitrary function $f(R, T)$ given in Eq. (\ref{eq6}). The matter's stress-energy tensor is elaborated as \cite{ref107a}:

\begin{eqnarray}
\label{eq7}
T_{\mu \nu}=-\frac{2}{\sqrt{-g}}\frac{\delta({\sqrt{-g}L_{m}})}{\delta g^{\mu \nu}} ,
\end{eqnarray}

Three exact stipulations of the functional anatomy of $f (R, T)$ have been considered in \cite{ref107} as
\begin{displaymath}
f(R,T) = {
\begin{array}{lr}
2f(T) + R, f(R,T) =
f_{2}(T) + f_{1}(R), f(R,T) =
f_{2}(R)f_{3}(T) + f_{1}(R)
\end{array}}.
\end{displaymath}

 One can obtain various theoretical models for each choice of $f$ because of $f(R, T)$ gravity's field equations also depend on the matter's field physical nature.\\
 
In the present contribution, we reflect the cosmological outcomes of the first class for which $f (R, T ) = 2f(T ) + R$. The field equation for this case by varying the action with $g_{\mu \nu}$ given by Eq. (\ref{eq6}), is   obtained as 
%\begin{eqnarray}
%\label{eq13} 
%R_{\mu \nu}-\frac{1}{2} R g_{\mu \nu} = 8\pi T_{\mu \nu} - 2 f^{'}_{T}(T)T_{\mu \nu}- 2 f^{'}_{T}(T)\Theta_{\mu\nu} + f(T) g_{\mu \nu},
%\end{eqnarray}

%Perfect fluid is the object source, $ \Theta_{\mu \nu} = - 2T_{\mu\nu} - pg_{\mu\nu}$, so the field equation tranforms to
%\begin{eqnarray}
%\label{eq14} 
%R_{\mu \nu}-\frac{1}{2} R g_{\mu \nu} = 8\pi T_{\mu \nu} + 2 f^{'}_{T}(T)T_{\mu \nu}+
%g_{\mu \nu}[ 2 p f^{'}_{T}(T) + f(T)] .
%\end{eqnarray}
%The field equations of gravitation in the plight of dust $p=0$ can be taken as:
\begin{eqnarray}
\label{eq15} 
R_{\mu \nu}-\frac{1}{2} R g_{\mu \nu} = 8\pi T_{\mu \nu} + 2 f^{'}_{T}(T)T_{\mu \nu}+f(T) g_{\mu \nu},
\end{eqnarray}
here the prime expresses a derivative concerning the argument. The authors proposed these field equations to resolve the cosmological constant enigma  \cite{ref108}. By taking $f(T )$ as the function so that $\lambda T = f(T )$, where constant is $\lambda$  and allowing a dust Universe $(\rho = T$,  $p = 0)$, the simplest model of cosmology can be achieved. 

\section{The cosmological model}

By appropriating that the metric of the Universe is provided by the flat FRW metric,
\begin{eqnarray}
\label{eq16}
ds^{2} = (dx^{2}+dy^{2}+dz^{2}) a^{2}(t) -  dt^{2}.
\end{eqnarray}
The field equations of $f(R, T)$ gravity for the metric given by Eq. (\ref{eq16})  are defined as
\begin{eqnarray}
\label{eq17}
3\frac{\dot{a}^{2}}{a^{2}} = \rho_{\Lambda} (3\lambda + 8 \pi ),
\end{eqnarray}
\begin{eqnarray}
\label{eq18}
2 \frac{\ddot{a}}{a} + \frac{\dot{a}^{2}}{a^{2}} = \lambda \rho_{\Lambda}.
\end{eqnarray}

 Hence, this model of $f(R, T )$ theory is comparable to a model of cosmology with an effective cosmological constant $\Lambda_{eff} \propto H^{2}$, where $H=\frac{\dot{a}}{a}$ is the Hubble parameter \cite{ref108}. Interestingly for the choice of $f(R, T )$, the form $G_{eff}=G\pm 2f^{'}(T)$ of gravitational coupling becomes time-dependent and an effective coupling. Displacing G through working on the parameter of the gravitational coupling, the interaction of gravitation among curvature and matter is modified by the term $2f(T )$ in the action of gravitation.  Hence, the field equation reduces to a single equation of $H$ \cite{ref107}, as

\begin{eqnarray}
\label{eq19}
2\dot{H}+3\frac{8\pi +2\lambda}{8\pi+3\lambda} H^{2} = 0,
\end{eqnarray}
general solution can be made as
\begin{eqnarray}
\label{eq20}
H(t)=\frac{2(2\lambda + 8\pi)}{3(3\lambda +8\pi)}\frac{1}{t}.
\end{eqnarray}
The scale factor unfolds agreeing to $t^{\beta} = a(t)$, where $\beta=\frac{2(8\pi +2\lambda)}{3(8\pi+3\lambda)}$ and provides the power law expansion $a\propto t^{\beta}$. Since, we obtain the solution for the case of  dark energy dominated Universe, the scale factor gives a matter dominated conduct for $\beta=2/3$, showing that the derived model avoids conflict with the coincidence problem. The similar expression for the Hubble parameter $H$ was found in \cite{ref93}, while proposing the GO cutoff, for the small  $t$ limit in \cite{refN2}, and using the future event horizon cut off in \cite{ref104}. From the equation of conservation, on the other side,
\begin{eqnarray}
\label{eq21}
\frac{\partial\rho_{\Lambda}}{\partial t}+ (\rho_{\Lambda}+p_{\Lambda}) ~3H=0,
\end{eqnarray}
For the holographic energy practicing pressure densities $p_{\Lambda} = \omega_{\Lambda} \rho_{\Lambda}$ and EoS (equation of state) of a barotropic, we found an illustration for the EoS parameter $\omega_{\Lambda}$ as
\begin{eqnarray}
\label{eq22}
\omega_{\Lambda}=-1+\frac{(2\delta-4)\dot{H}}{3 H^{2}}.
\end{eqnarray}
%%%%%%%%%%%%%%%%%%%%%%%%%%%%%%%% Figure 1 %%%%%%%%%%%%%%%%%%%%%%%%%%%%%%%%%%%%%%%%%%%%%%%%%%%%%%%%
%\begin{figure}[htbp]
%	\centering
%	\includegraphics[width=10cm,height=10cm,angle=0]{fig1.eps}
%	\caption{The performance of the EoS (equation of state) parameter $\omega_{\Lambda}$ versus redshift (z).} 
%	
%\end{figure}

Using Eq. (\ref{eq20}) in Eq. (\ref{eq22}), we can find
\begin{eqnarray}
\label{eq23}
\omega_{\Lambda}=-1-(\delta-2)\bigg[\frac{8\pi +2\lambda}{8\pi+3\lambda}\bigg],
\end{eqnarray}
Here, by use of $\omega_{\Lambda} > -1$, which is the allowed range for quintessence \cite{ref21},  we can obtain the constraints of the parameters $\delta$ and $\lambda$. If $ \omega_{\Lambda}>-1$, the rate of expansion of the Universe accelerates always \cite{ref21}. In Eq. (\ref{eq23}), the EoS $\omega_{\Lambda}$ is explained in terms of the constants $\lambda$ and $\delta$. The constants $\lambda$ and $\delta$ must justify the restrictions followed from Eq. (\ref{eq23}), to get accelerated expansion. If $-6\pi<\lambda < -4\pi$ and $0<\delta < 2$ then we consider the $\omega_{\Lambda} > -1$ phase and for $-6\pi<\lambda < -4\pi$ and $\delta > 2$ the density of Tsallis holographic shows the evolution's phase as a phantom with $\omega_{\Lambda} < -1$. While, for $\lambda = -4\pi$ or $\delta = 2$, the EoS  $\omega_{\Lambda} = -1$, mimics the cosmological constant.\\

\begin{figure}[htbp]
	\centering
	\includegraphics[width=14cm,height=8cm,angle=0]{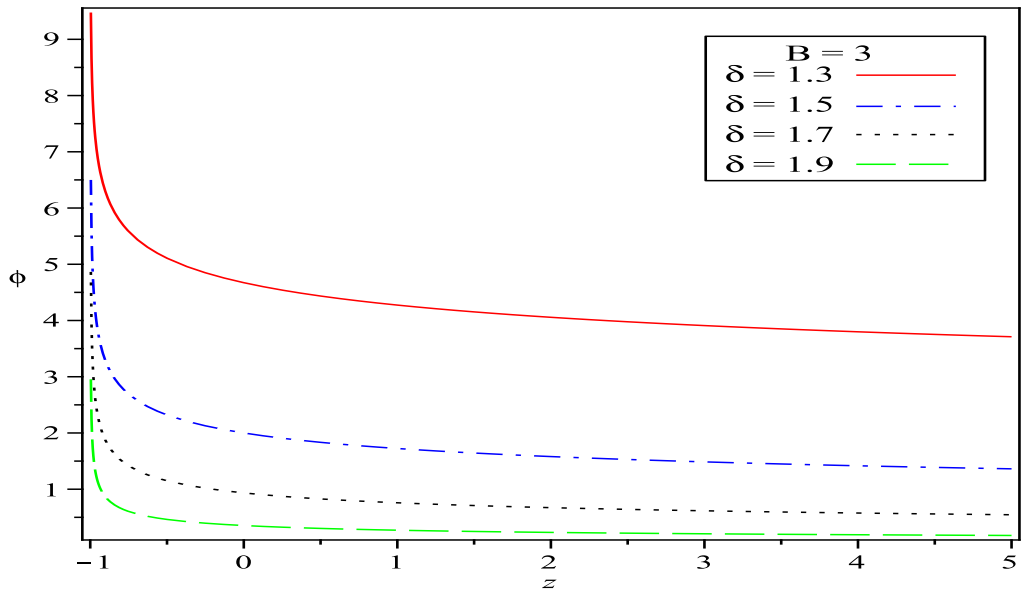}
	\caption{The behaviour of Tsallis holographic quintessence scalar field $(\phi)$ against redshift ($z$) for THDE 1 and $\lambda = -5\pi$.} 
	
\end{figure}

It is necessary to specify that differentiating the frame of Tsallis HDE with the standard HDE, Tsallis HDE provides more extreme behavior and quantifies at the presence of the novel parameter $\delta$. One can obtain the sub-case of the standard HDE framework because of its agreeable foundation, which is distinct for $\delta$ = 1 \cite{refN3}. One should perform many more researches for the description of nature, before the circumstances considered as a profitable contender. Importantly, for compelling the model parameters, one should strike out shared empiric research using large scale structure and data from CMB at each confusion background and levels. To classify the Universal highlights of the framework in modern times, we ought to perform the analysis of phase space in detail, the underlying conditions independently, and the specific evolving \cite{refN3}.\\

Approaching, Universe's accelerated expansion form one of the current works \cite{refN4}, the researcher recognized the legitimacy of the holographic principle. Based on the entropy of Tsallis, the researcher explained the highlights of the DE model, which says of the BG entropy's non-additive generalization that needed to use in the scientific summary of some non extensive frames, such as gravitation. The researcher reflected on the IR cutoff for the future event horizon in this research. This choice made by the researcher has allowed him to include standard thermodynamics and standard HDE as sub-classes and to avoid the cosmological variations. The researcher has divided his work into two distinct THDE conditions: ~the primary model recovers standard HDE with a fixed value of $B=3$ in the limit of $\delta$ = 1, while they left $B$ as an independent parameter for the secondary model. Subsequently, they tried to make a correlation among the latest observation of cosmology and the observation of earlier specified hypothetical models \cite{refN4}. \\

In  \cite{refN4}, the researcher had practiced the data from the GRF (growth rate factor) matter fluctuations and combined it with the data from type SN (Ia Supernovae) and OHD (observational Hubble data) to restrain Tsallis DE model as IR cutoff for the future event horizon and obtained $ 68\%$ $(95\%)$ confidence level and possibilities of the parameter of cosmology assuming from the study of MCMC (Markov Chain Monte Carlo) with independent $B$ and fixed $B=3$ of THDE models. $\delta$ the nonextensive Tsallis parameter, and $B$  the model parameter are constraints according to the research of the observational data for Tsallis DE model as  $\delta = 0.939^{+0.053(0.107)}_{-0.054(0.101)}$ and $B=3$, and  $B =  2.864^{+0.741(2.405)}_{-1.454(1.821)}$  and $\delta = 0.941^{+0.053(0.104)}_{-0.054(0.101)}$ \cite{refN4}.\\

In \cite{refN5}, they have examined briefly about the observational limitations of THDE model as IR cutoff with Hubble horizon. They applied various observational data of cosmology, BAO ( Baryon acoustic oscillation), data of the Gamma-Ray Burst, the Pantheon SN (type Ia supernovae), the confined value of $H_{0}$ (Hubble constant) and CMB (cosmic microwave background) for restraining independent parameter. $B=3$ and $\delta = 1.871_{-0.441}^{+0.190}$ are most commendable results in \cite{refN5}. \\

As restrained in \cite{refN4, refN5}, we used the model parameter's value and expressed it as THDE 1 and THDE 2 in my research work.\\

{\bf THDE 1}\\

For the current contribution, we present the model parameter's value as $\delta$ = 1.3, 1.5, 1.7, 1.9 and $B$ = 3 in the primary framework.\\

{\bf THDE 2}\\

For the current contribution, we present the model parameter's value as $\delta$ = 1.3, 1.5, 1.7, 1.9 and $B$ = 2.8 in the secondary framework.\\ 

It is also proposed in \cite{refN3}, that there are extensive limits for $\delta$ and $B$ which can produce aspired results. Comparing the results with those
	of \cite{refN3,refN4,refN5}, we can easily see that  the
	values of $\delta$ and $B$ used in this work place within the allowed
	interval of $\delta$ and $B$ reported by \cite{refN3,refN4,refN5}.
\begin{figure}[htbp]
	\centering
	\includegraphics[width=16cm,height=8cm,angle=0]{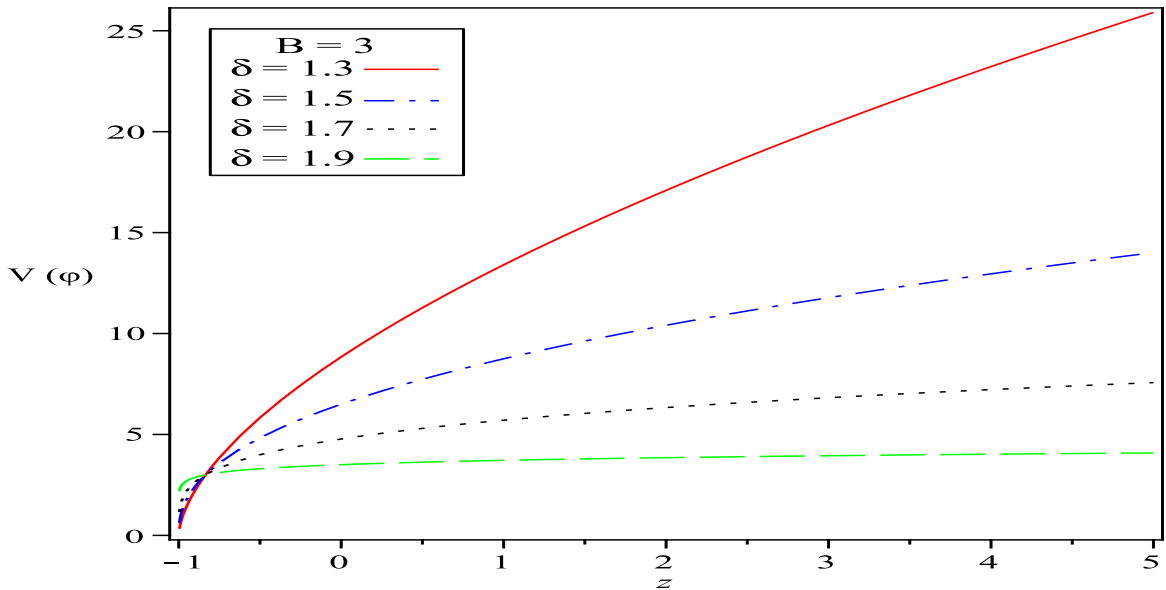}
	\caption{The behaviour of Tsallis holographic DE quintessence potential against redshift ($z$) for THDE 1 and $\lambda = -5\pi$.} 
\end{figure}

\section{Reconstruction of quintessence model for THDE}

In this segment, we shall investigate the correspondence between the quintessence and the THDE in a flat $f(R, T)$ Universe for the case THDE 1 and THDE 2. To establish the  correspondence between the quintessence and the THDE model, we compare the  equation of state for the quintessence field with the THDE model given by Eq. (\ref{eq23}), and also equate the THDE model  energy density  given by Eq. (\ref{eq5}) with the corresponding energy density of quintessence scalar field model. Furthermore, we shall reconstruct the dynamics and the potentials for the quintessence model. We can produce the corresponding outcomes of the potentials and the quintessence scalar field for the both cases of  THDE model in $f(R, T)$ gravity.\\

\textbf{{\large THDE 1}}\\

 The pressure and the energy density of the quintessence field in FRW background \cite{ref21}

\begin{eqnarray}
\label{eq24}
\rho_{\phi} = V(\phi) + \frac{1}{2}\dot{\phi}^{2},
\end{eqnarray}
\begin{eqnarray}
p_{\phi} = - V(\phi) + \frac{1}{2}\dot{\phi}^{2} 
\end{eqnarray}
For the scalar field, the equation of state (EoS) is given as
\begin{eqnarray}
\label{eq25}
\omega_{\phi} = \frac{-2V(\phi) + \dot{\phi}^{2}}{2V(\phi) + \dot{\phi}^{2}},
\end{eqnarray}
which on comparing with the Tsallis holographic EoS parameter Eq. (\ref{eq23}), provides the equation
\begin{eqnarray}
\label{eq26}
 \frac{-2V(\phi) + \dot{\phi}^{2}}{2V(\phi) + \dot{\phi}^{2}} = -1-(\delta-2)\bigg[\frac{8\pi +2\lambda}{8\pi+3\lambda}\bigg]
\end{eqnarray}
which together with the equation
\begin{eqnarray}
\label{eq27}
\rho_{\phi}~ = ~ V(\phi) + \frac{1}{2}\dot{\phi}^{2}~ =~ BH^{-2\delta+4}
\end{eqnarray}
%%%%%%%%%%%%%%%%%%%%%%%%%%%%%%%% Figure 2 %%%%%%%%%%%%%%%%%%%%%%%%%%%%%%%%%%%%%%%%%%%%%%%%%%%%%%%%

to lead the exact expressions for the potential and the scalar field, this equation can be solved, specifically,

\begin{eqnarray}
\label{eq28}
\phi = \pm \frac{\sqrt{B(2-\delta)}}{\delta-1}\bigg(\frac{2}{3}\bigg)^{-\delta+2} \bigg[\frac{8\pi+3\lambda}{8\pi +2\lambda}\bigg]^{\frac{-2\delta+3}{2}} t^{\delta-1}. 
\end{eqnarray}

The expression given by Eq. (\ref{eq28}) represents the scalar fied and is plotted in Fig. 1.
Here, we have taken  the +ve sign in Eq. (\ref{eq28}). We observe from this figure that with the increase in the red shift $z$, the scalar field $\phi$ decreases, displays constant at the high red shift and $\phi$ becomes finite at future. The similar behavior was observed for the scalar field $\phi$ in \cite{refN6} for the $k$-essence scalar field \cite{ref109}. Using Eq. (\ref{eq20}) in Eq. (\ref{eq27}) and then solving with Eq. (\ref{eq26}), we can get the expression for the potential in terms of scalar field $\phi$ as

%%%%%%%%%%%%%%%%%%%%%%%%%%%%%%%% Figure 3 %%%%%%%%%%%%%%%%%%%%%%%%%%%%%%%%%%%%%%%%%%%%%%%%%%%%%%%%

\begin{eqnarray}
\label{eq29}
V(\phi) = B ~ \bigg(\frac{2}{3}\bigg)^{-2\delta+4} ~  
 \bigg[\frac{\lambda + 4 \delta \pi + \delta \lambda}{8 \pi + 3 \lambda}\bigg]~ \bigg[\frac{8\pi+3\lambda}{8\pi +2\lambda}\bigg]^{-2\delta+4}~ A\\ \nonumber
 A = \Bigg[~ \frac{\sqrt{B(2-\delta)}}{\phi(\delta-1)}~ \bigg(\frac{2}{3}\bigg)^{-\delta+2}  ~ \bigg[\frac{8\pi+3\lambda}{8\pi +2\lambda}\bigg]^{\frac{-2\delta+3}{2}}  \Bigg]^\frac{-2\delta+4}{\delta-1}
 \end{eqnarray}
 
 %%%%%%%%%%%%%%%%%%%%%%%%%%%%%%%% Figure 4 %%%%%%%%%%%%%%%%%%%%%%%%%%%%%%%%%%%%%%%%%%%%%%%%%%%%%%%%
 \begin{figure}[htbp]
 	\centering
 	\includegraphics[width=14cm,height=8cm,angle=0]{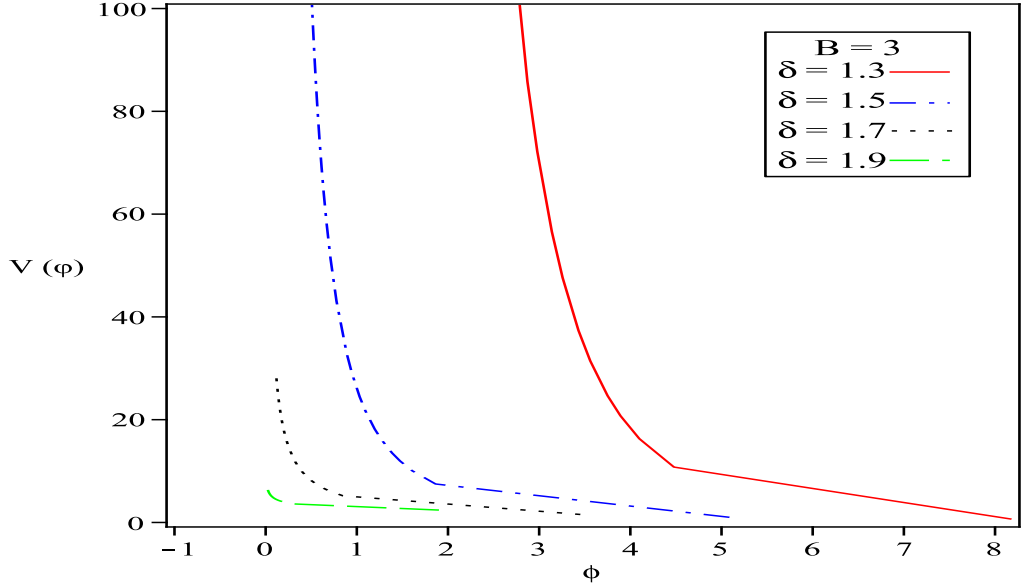}
 	\caption{The behavior of Tsallis holographic DE quintessence potential against the scalar field ($\phi$) for THDE 1 and $\lambda = -5\pi$.} 
 \end{figure}
 
 We have plotted the quintessence potential with red shift $z$ in Fig. 2, for the THDE 1 model.
We observe from Fig. 2 that  the potential decreases as we proceed from past to future. The similar behavior for the quintessence potential has been observed in \cite{ref105a} for the holographic quintessence model, and for the holographic tachyon model in \cite{ref102}. We have also plotted the quintessence potential versus the scalar field as exhibited in Fig. 3.

\begin{figure}[htbp]
	\centering
	\includegraphics[width=14cm,height=8cm,angle=0]{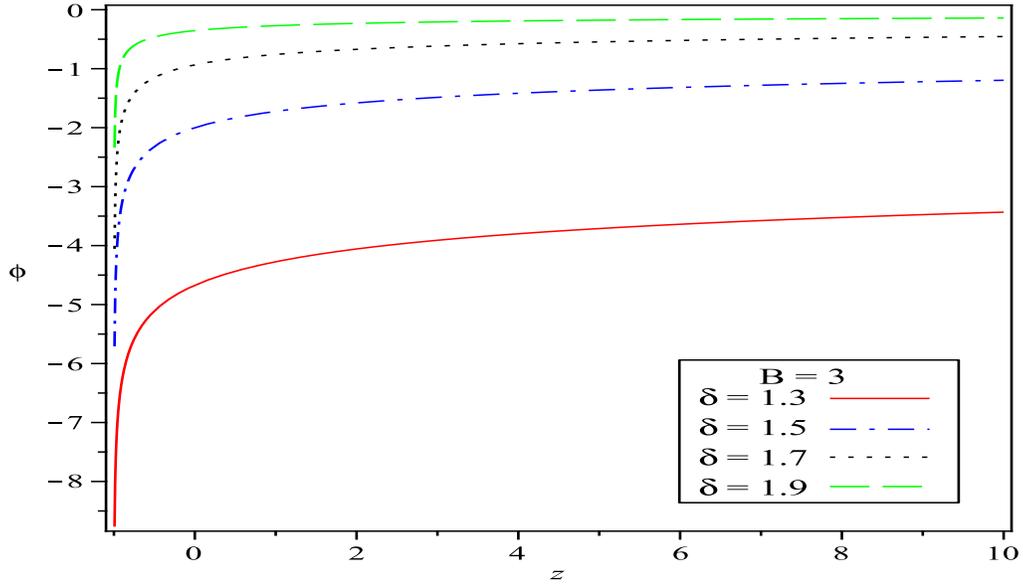}
	\caption{ The behaviour of Tsallis holographic DE quintessence scalar field against redshift ($z$) for THDE 1 and $\lambda = -5\pi$. Here, the negative sign is taken in Eq. (\ref{eq28}). } 
\end{figure}

It is clearly seen from Fig. 3, that  the potential decreases as the scalar field increases for the THDE 1 model. The same behavior was shown in \cite{ref109a} for the quintessence potentials. If we take the negative sign in  Eq. (\ref{eq28}), we can too evaluate the potential and at present. In Fig. 4, we can see the shift of the scalar field, because of this result, it can not modify the shape of the potential, and also can not influence the cosmological evolution of the Universe. In this case, we can see the appearance of the quintessence field in Fig. 5, by replacing the sign of the scalar field. One can see from Fig. 5, the potential is more steep in the past, approaching to be flat at future implies that the potential is rolled down by the quintessence field as the Universe expand. The same conclusion was observed in \cite{ref105a} for the quintessence potential. Therefore, it was proposed in \cite{ref109a} that the potentials are of a runway type, and decrease as the Universe expands. \\

 %%%%%%%%%%%%%%%%%%%%%%%%%%%%%%%% Figure 6 %%%%%%%%%%%%%%%%%%%%%%%%%%%%%%%%%%%%%%%%%%%%%%%%%%%%%%%%
 \begin{figure}[htbp]
 	\centering
 	\includegraphics[width=14cm,height=8cm,angle=0]{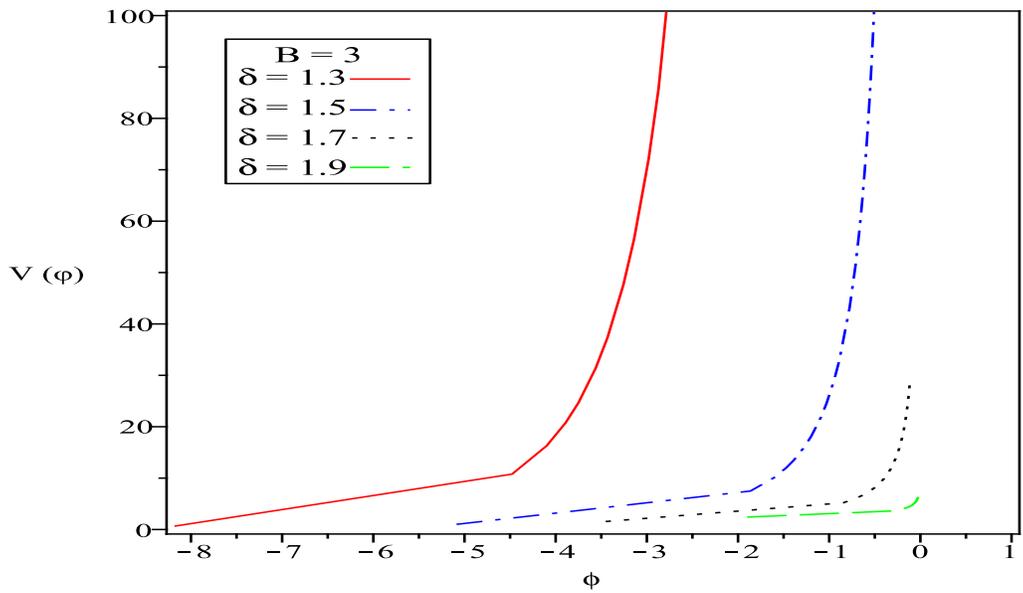}
 	\caption{ The behavior of the Tsallis holographic DE quintessence potential against the scalar field ($\phi$) for the THDE 1 and $\lambda = -5\pi$. Here, the negative sign is taken in Eq. (\ref{eq28}).} 
 \end{figure}

\textbf{{\large THDE 2}}\\

Copying the same style as $B  = 3$, we will receive the shape of the potential and the scalar field that recognizes the accelerated expansion for the case $B = 2.8$. In this case, Fig. 6 imitates the scalar field in terms of red shift, Fig. 7 draws the potential in terms of red shift, and Fig. 8 portrays the potential in terms of the scalar field for the quintessence field.\\

%%%%%%%%%%%%%%%%%%%%%%%%%%%%%%%% Figure 7 %%%%%%%%%%%%%%%%%%%%%%%%%%%%%%%%%%%%%%%%%%%%%%%%%%%%%%%%
\begin{figure}[htbp]
	\centering
	\includegraphics[width=14cm,height=8cm,angle=0]{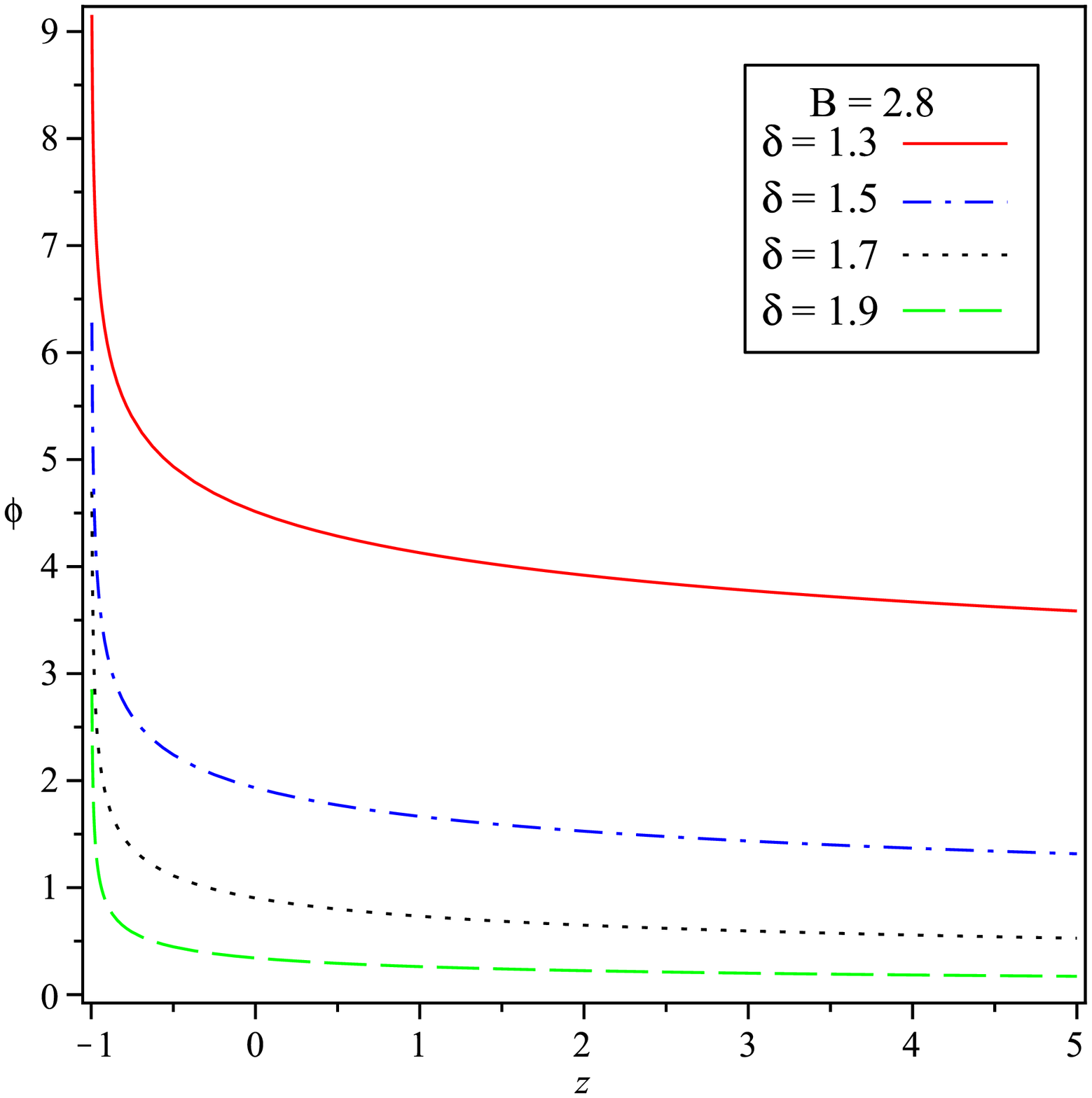}
	\caption{ The behaviour of the Tsallis holographic DE quintessence scalar field against redshift ($z$) for the THDE 2 and $\lambda = -5\pi$.} 
	
\end{figure}
%%%%%%%%%%%%%%%%%%%%%%%%%%%%%%%% Figure 8 %%%%%%%%%%%%%%%%%%%%%%%%%%%%%%%%%%%%%%%%%%%%%%%%%%%%%%%%
\begin{figure}[htbp]
	\centering
	\includegraphics[width=14cm,height=8cm,angle=0]{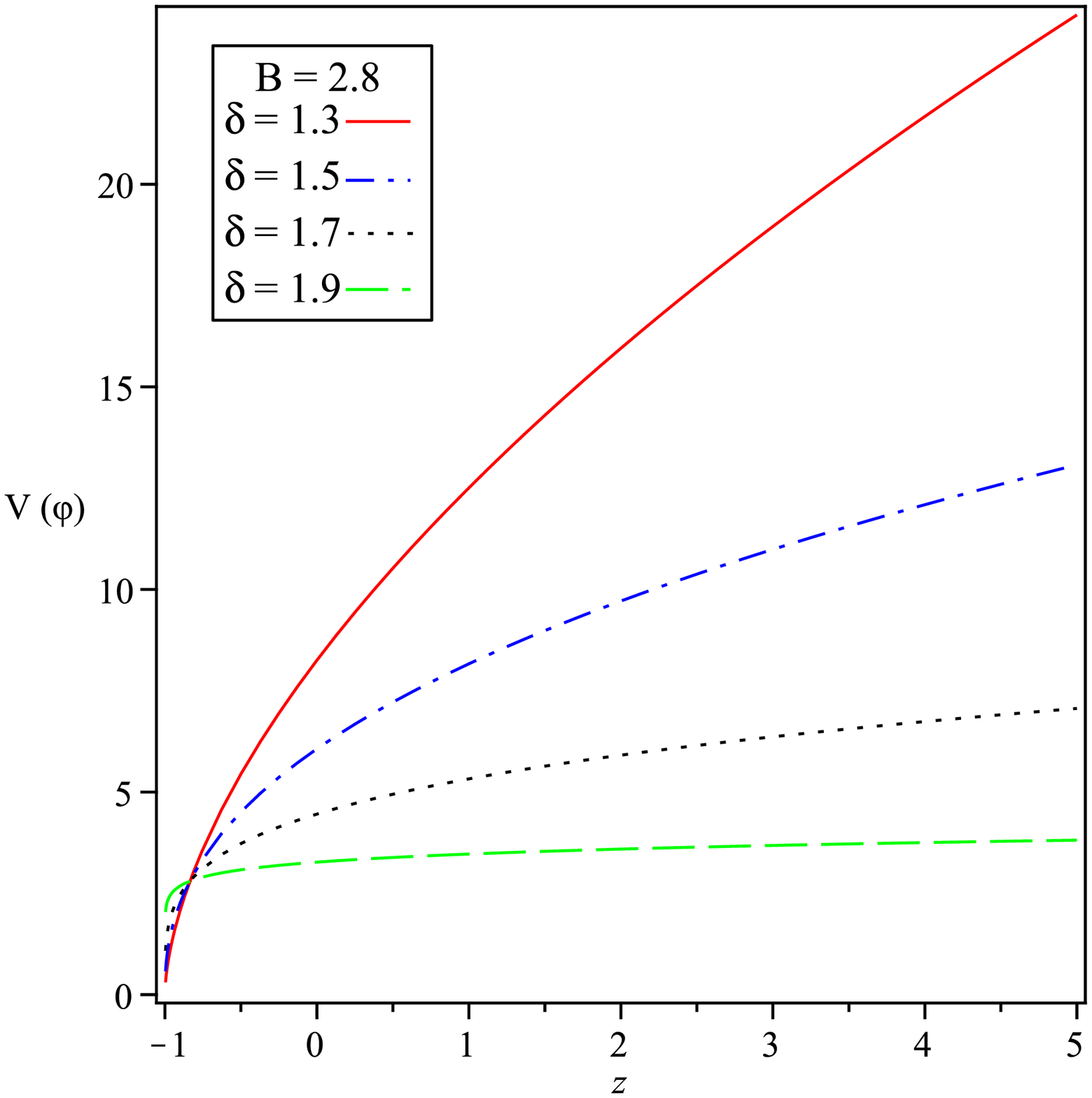}
	\caption{The behavior of Tsallis holographic DE quintessence potential against red shift ($z$) for THDE 2 and $\lambda = -5\pi$. } 
	
\end{figure}
%%%%%%%%%%%%%%%%%%%%%%%%%%%%%%%% Figure 9 %%%%%%%%%%%%%%%%%%%%%%%%%%%%%%%%%%%%%%%%%%%%%%%%%%%%%%%%
\begin{figure}[htbp]
	\centering
	\includegraphics[width=14cm,height=8cm,angle=0]{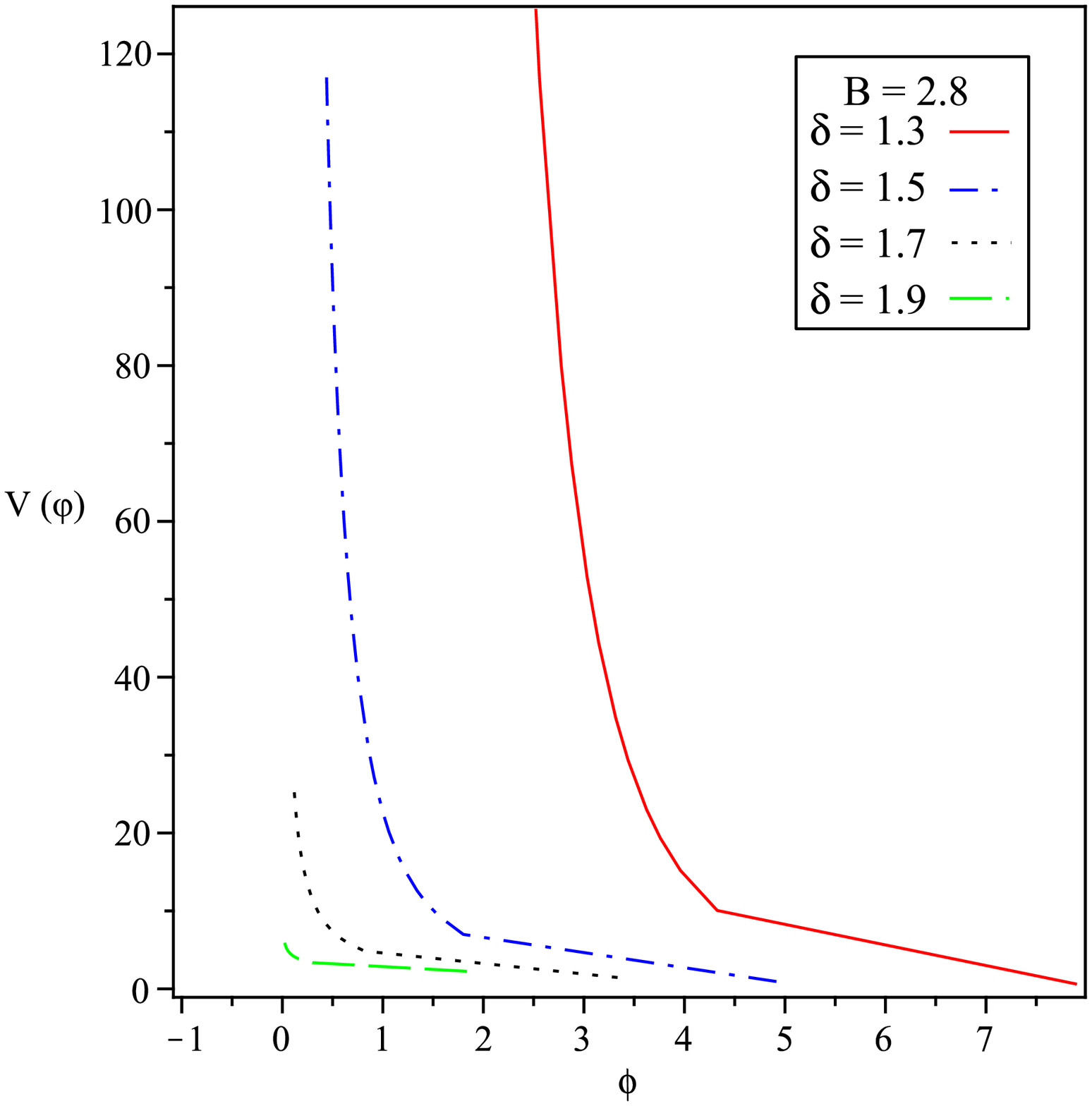}
	\caption{The behaviour of the Tsallis holographic DE quintessence potential against the scalar field ($\phi$) for the THDE 2 and $\lambda = -5\pi$. } 
	
\end{figure}

As in Eq. (\ref{eq28}), we turn the negative sign of the scalar field, we can see the shift of the scalar field in Fig. 9, and the shift of the potential versus the scalar field in Fig. 10 for the quintessence field. 
It is remarked that the dynamics of the quintessence behave in the same manner as in the $B=3$ case.\\

%%%%%%%%%%%%%%%%%%%%%%%%%%%%%%%% Figure 10 %%%%%%%%%%%%%%%%%%%%%%%%%%%%%%%%%%%%%%%%%%%%%%%%%%%%%%%%
\begin{figure}[htbp]
	\centering
	\includegraphics[width=14cm,height=8cm,angle=0]{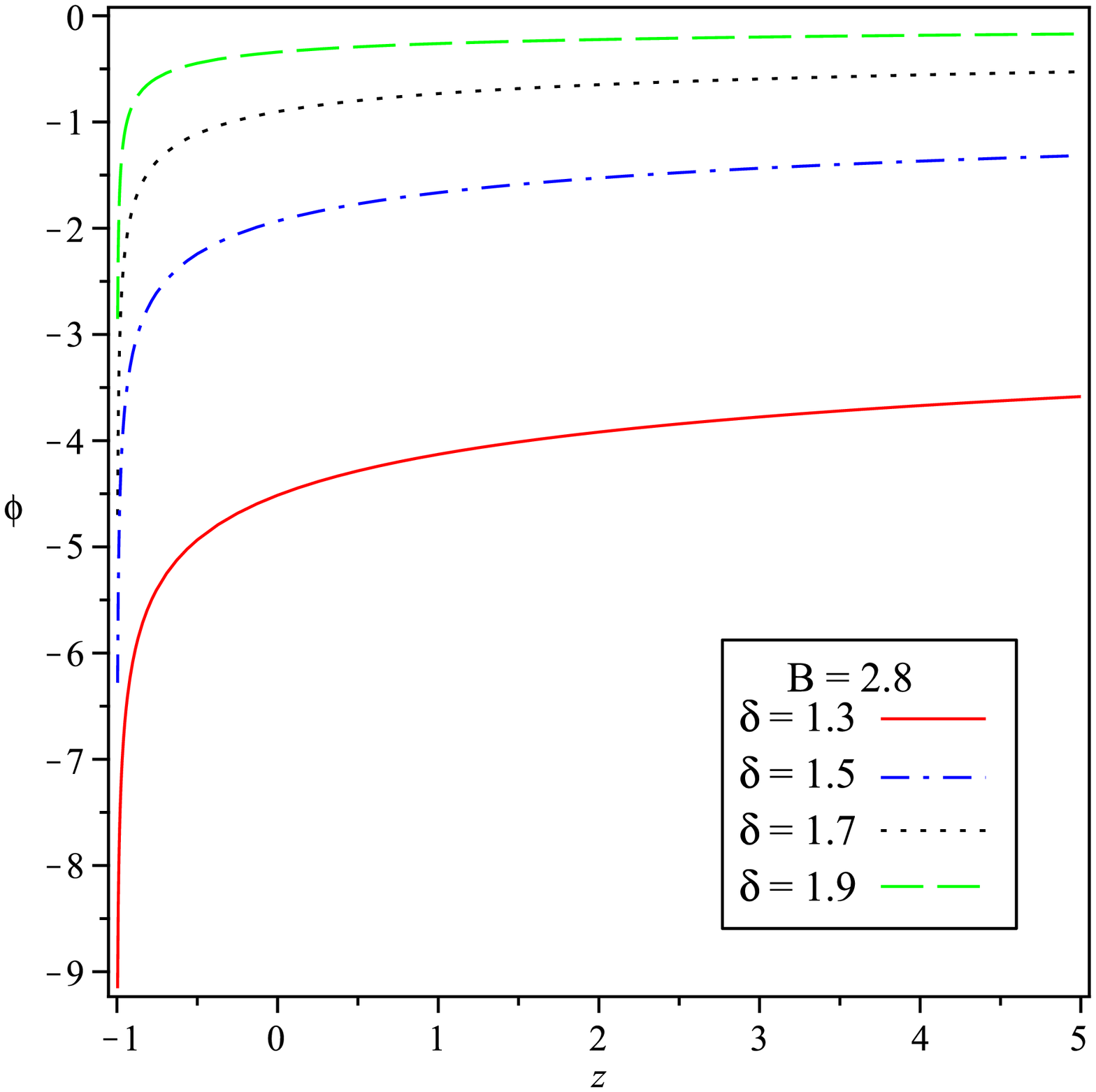}
	\caption{The behavior of the Tsallis holographic DE quintessence scalar field against the red shift ($z$) for the THDE 2 and $\lambda = -5\pi$. Here, the negative sign is taken in Eq. (\ref{eq28}).} 
	
\end{figure}

%%%%%%%%%%%%%%%%%%%%%%%%%%%%%%%% Figure 11 %%%%%%%%%%%%%%%%%%%%%%%%%%%%%%%%%%%%%%%%%%%%%%%%%%%%%%%%
\begin{figure}[htbp]
	\centering
	\includegraphics[width=14cm,height=8cm,angle=0]{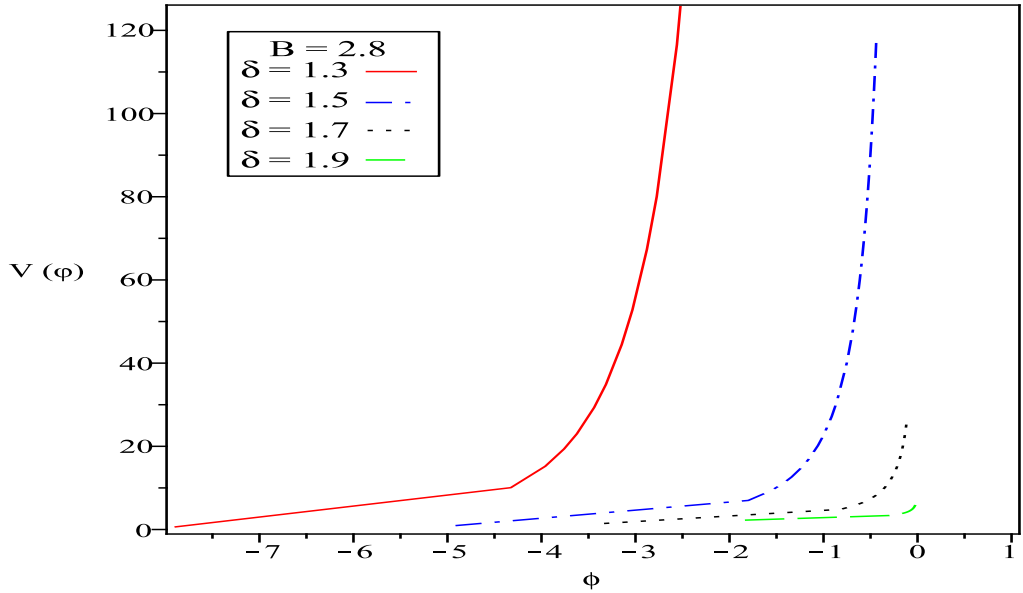}
	\caption{ The behavior of the Tsallis holographic DE quintessence potential against the scalar field ($\phi$) for the THDE 2 and $\lambda = -5\pi$. Here, the negative sign is taken in Eq. (\ref{eq28}).} 
	
\end{figure}

\section{ The correspondence with the Swampland criteria  of quintessence phase for THDE model} 

In modern times a new interest is developed to connect applicable models of cosmology with gravity's quantum theory, which can easily differentiate between the effective field models related to the nominal string landscape. Recently Swampland criteria are introduced to choose field potentials effectively. Precisely, the excitement aroused because accurate solutions of de Sitter with a positive constant of cosmology is not fit with the string landscape, which means it cannot link at the arrangement of high energy with some fundamental theory. Lately, the researcher investigated the connection of the Swampland conjecture and the viability of the $f(R)$ theory \cite{ref123}. Several philosophical contexts reviewed, the DE phase as we have explained in the introduction of the paper, for example, quintessence models, modified gravity, and so on, and the hurdle to get the compatibility with the observational data represented.\\

Newly, \cite{ref110, ref111} derived the swampland criteria of the string theory constrained with the quintessence model of DE. Approaching the Swampland criteria in the connection of DE's quintessence models, many papers reported after the work in \cite{ref110, ref111}, for example, \cite{ref112, ref113, ref114, ref115, ref116, ref117, ref118, ref119, ref120, ref121, ref122, ref123}. The Swampland criteria restrain and rules on those field theories which are conflicting with quantum gravity \cite{ref124}. Various researchers submitted the work satisfying the low energy effective theory  \cite{ref125} for the scalar field in the Swampland conjecture,

\begin{eqnarray}
\label{eq30}
Conjecure \hspace{0.5cm} 1 :\hspace{2cm} \mid\Delta\phi\mid \leq M_{P}d_{1},
\end{eqnarray}

\begin{eqnarray}
\label{eq31}
Conjecure \hspace{0.5cm} 2 :\hspace{2cm} \mid\Delta V\mid \geq \dfrac{d_{2}V}{M_{P}}.
\end{eqnarray}

If the conjecture has UV (ultraviolet) completion compatible with quantum gravity, then over a specific limit of the scalar field the +ve constants $d_{2}$ and $d_{1}$ are of Ist order, and the conjecture lies in the Swampland criteria, and also the potential of scalar is too flat.\\ 

In the analysis of DE, the quintessence models plays an important role, therefore in this work, we have examined the performance of the THDE model in the connection of  Swampland conjecture for quintessence case. Eq. (\ref{eq28}) shows the scalar field $\phi$ that is associated with swampland criteria $1$, and  Eq. (\ref{eq29}) shows the scalar potential $V$ that is associated with swampland criteria $2$. The scalar field $\phi$ in the terms of red shift $(z)$ and the potential field $V$ in the terms of the scalar field $\phi$ are framed for  both the models THDE 1 and THDE 2 shown in Fig. 1-10. It is shown in these Figs. that the potential field is +ive while the scalar field is not at minimum, as shown in \cite{ref112} for the models of quintessence.

\section{Discussion}

In this work, we have brought up a comprehensive analysis for the cosmological evolution of the Tsallis holographic quintessence model of DE in the framework $f(R, T)$ gravity with the infrared cut off as Hubble horizon. In this model, we have three parameters $\delta$, $\lambda$, and $B$. The parameter $\delta$ performs a major role in the Tsallis holographic evolution of the Universe. We have investigated $\omega_{\Lambda} > -1$ phase, if $0<\delta < 2$  and $-6\pi<\lambda < -4\pi$ and the Tsallis holographic EoS represents the phase of evolution as a phantom with $\omega_{\Lambda} < -1$ for $\delta > 2$ and $-6\pi < \lambda < -4\pi$. While, the EoS of the THDE copies the cosmological constant  $\omega_{\Lambda} = -1$ for $\delta = 2$ or $\lambda = -4\pi$.
We have reconstructed the potential and the scalar field using the conformity with the Tsallis holographic density and with the equation of state parameter in the state of quintessence field for both the cases of the THDE model i.e. the THDE 1 and the THDE 2 in the region  $\omega_{\Lambda}> -1$ which is the defined region for the quintessence era. From Figs. 1-5, we may observe the quintessence field dynamics for the THDE 1 model. Fig. 5, depicts that at the early times the potential is steep and turning flat at the present and at future. Hence, based on the Hubble horizon cut off, the kinetic term is continuously decreasing for the Tsallis holographic quintessence model as witnessed in \cite{ref126,ref105a} for the negative sign  in the expression of Eq. (\ref{eq28}). For the positive sign in the expression of Eq. (\ref{eq28}), Fig. 3 shows the appearance of the potential and relates this work to \cite{ref109a}.  We have achieved the same result for the reconstruction of Tsallis quintessence potential as achieved in \cite{ref127} and \cite{ref128}. In both instances, Fig. 3 and Fig. 5 with the expansion in the Universe, the potential decreases as presented in Fig. 2. The reconstruction for the case of the THDE 2,  Figs. 6-10 gives the comparable result as the case THDE 1. We additionally examine the swampland conjecture for the Tsallis HDE model and the outcome is fascinating. Particularly observed that the potential ($V (\phi)$) is asymptotically flat as Universe expands.\\

According to the results presented in the literature, the reconstruction is beneficial in representing the main characteristics of the potential for the quintessence model. Although the Tsallis holographic model, which depends on the parameters, lifts the reconstruction unambiguously and immediately, which is shown phenomenological, still one should understand and review the hypothetical root of Tsallis holographic density. Nevertheless, we appropriately understand through the obtained result about the interesting problems of the cosmology.\\

Based on the model, to explain Tsallis holographic dark energy the reconstruction of the phantom, tachyon, $k$-essence and dilation scalar field can show detailed or meaningful effects on the evolution of the Universe. In the future paper, we shall spout these studies. As witnessing the Universe's future, my work in this paper contributes a different and new, reasonably pleasing opportunity to understand the nature of the THDE in the modified gravity.

\end{document}